\documentclass[aps,prl,reprint,superscriptaddress,letterpaper]{revtex4-1}
\usepackage[english]{babel}
\usepackage{ucs}
\usepackage[utf8x]{inputenc}
\usepackage{amsmath}
\usepackage{amsfonts}
\usepackage{amssymb}
\usepackage{graphicx}
\usepackage{bm}
\usepackage{bbm}
\usepackage{empheq}

\begin{document}

\title{A Radio Frequency Charge Parity Meter}

\author{M. D. Schroer}
\author{M. Jung}
\author{K. D. Petersson}
\affiliation{Department of Physics, Princeton University, Princeton, New Jersey 08544, USA}
\author{J. R. Petta}
\affiliation{Department of Physics, Princeton University, Princeton, New Jersey 08544, USA}
\affiliation{Princeton Institute for the Science and Technology of Materials (PRISM), Princeton University, Princeton,
New Jersey 08544, USA}

\date{\today}

\begin{abstract}
We demonstrate a total charge parity measurement by detecting the radio frequency signal that is reflected by a lumped element resonator coupled to a single InAs nanowire double quantum dot. The high frequency response of the circuit is used to probe the effects of the Pauli exclusion principle at interdot charge transitions. Even parity charge transitions show a striking magnetic field dependence that is due to a singlet-triplet transition, while odd parity transitions are relatively insensitive to magnetic field. The measured response agrees well with cavity input-output theory, allowing accurate measurements of the interdot tunnel coupling and the resonator-charge coupling rate $g_{\rm c}/2\pi\sim$ 17 MHz.
\end{abstract}


\maketitle
Double quantum dots (DQD) allow single electrons to be trapped in a fully tunable double well confinement potential \cite{Wiel2002} and have been used to study the crossover from weak to strong coupling in an artificial molecule \cite{Oosterkamp1998}, spin physics associated with the Pauli exclusion principle \cite{Ono2002,Johnson2005a}, and quantum control of both charge and spin states \cite{Hayashi2003,Petta2004,Petta2005,Hanson2007}. Traditional quantum dot experiments measure the response due to a low frequency ($f$ $<$ 500 Hz) electrical excitation, allowing the energy level spectrum to be probed in a manner that is analogous to optical spectroscopy of atomic systems \cite{Kouwenhoven1997}.

A number of novel phenomena have been studied in the high frequency response of mesoscopic systems, including a violation of Kirchhoff's laws at GHz frequencies \cite{Buttiker1993,Pretre1996,Gabelli2006,Mora2010,Petersson2010,Cottet2011}. In this regime, the electrical response can be characterized by a ``mesoscopic admittance", $Y(\omega) = R_{\rm eff}^{-1} + i \omega C_{\rm eff}$, where the effective resistance $R_{\rm eff}$ and capacitance $C_{\rm eff}$ are influenced by electronic tunneling and the density of states at the Fermi level \cite{Chorley2012}. Coupling the electric field of a microwave cavity to the dipole moment of a qubit has allowed the exploration of cavity quantum electrodynamics in a solid state environment, opening up new avenues of research in mesoscopic physics \cite{Wallraff2004,Delbecq2011,Frey2012,Petersson2012}. However, the interplay of spin and charge has not yet been examined in measurements of the mesoscopic admittance of DQD devices.

In this Letter, we probe spin-dependent effects in the radio frequency (rf) response of a DQD and develop a method for measuring the total charge parity of a DQD. We perform charge sensing by coupling a rf lumped element resonator to a single InAs DQD \cite{Petersson2010}. The signal reflected by the resonator is a sensitive probe of the charge state of the sample, allowing a determination of the absolute charge number \cite{Jung2012}. We show that the charge sensing signal at interdot charge transitions is sensitive to the total charge parity of the DQD at high magnetic fields, where the total charge parity refers to whether or not the total electron number is even or odd. The magnetic field, temperature, and tunnel coupling dependence of the reflected rf signal agrees well with cavity input-output theory, and gives a resonator-charge coupling rate of $g_{\rm c}/2\pi$ $\sim$ 17 MHz \cite{Walls1995,Clerk2010}.

\begin{figure}[t]
\begin{center}
		\includegraphics[width=\columnwidth]{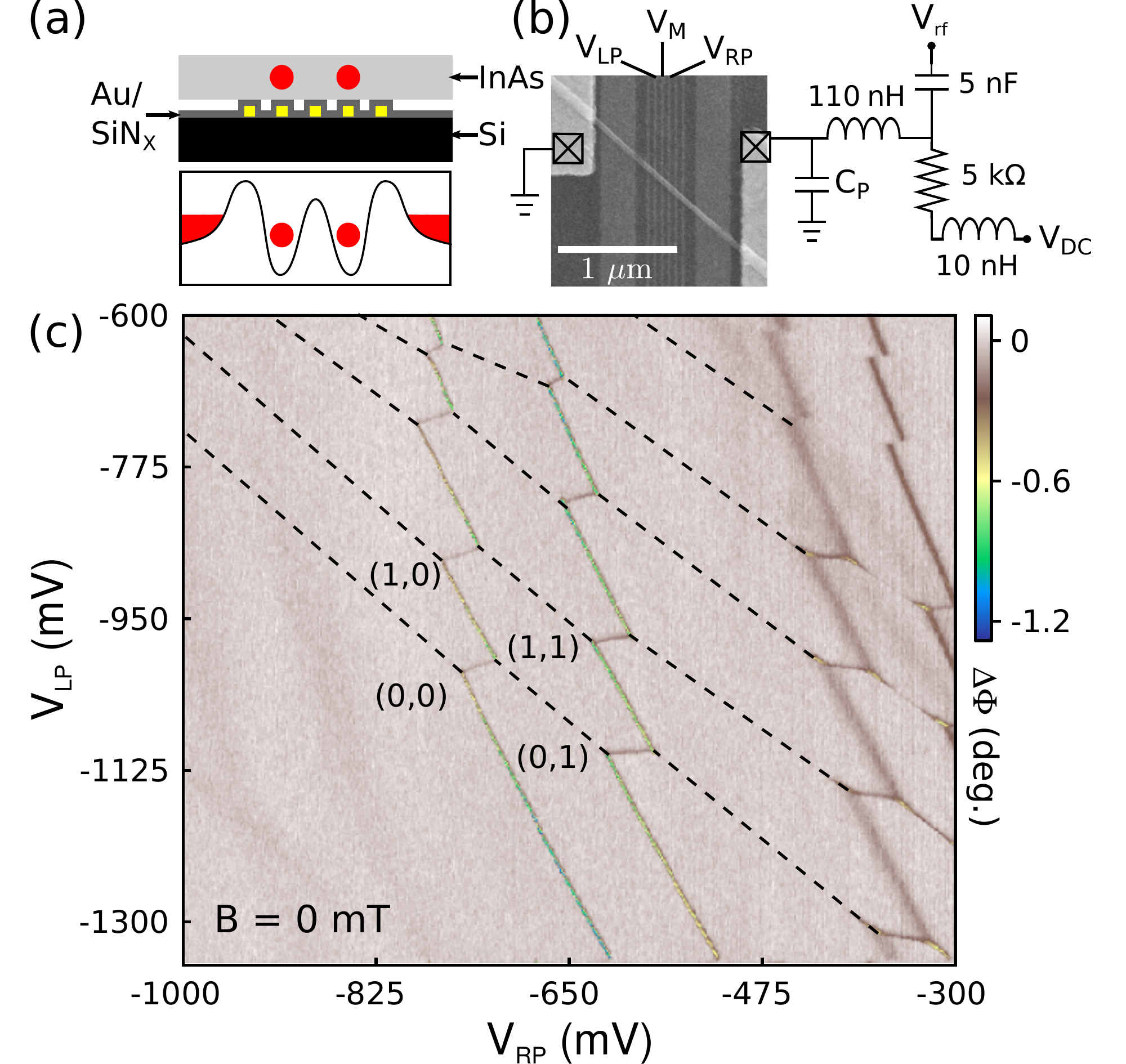}
\caption{\label{sense1} (Color online) (a) A DQD is defined along the length of a $\sim$50 nm diameter InAs nanowire using an array of bottom gate electrodes. (b) A SEM image of a typical device. The internal charge state of the DQD is probed by measuring the signal reflected by a lumped-element resonator that is coupled to the source contact of the nanowire. A bias tee allows for dc transport measurements. (c) The phase response of the reflected signal, $\Delta \Phi$, measured as a function of the gate voltages $V_{\rm LP}$ and $V_{\rm RP}$, determines the DQD charge stability diagram. The dashed lines are included as guides to the eye.}
\end{center}	
\end{figure}

\begin{figure*}[t]
\begin{center}
		\includegraphics[width=2\columnwidth]{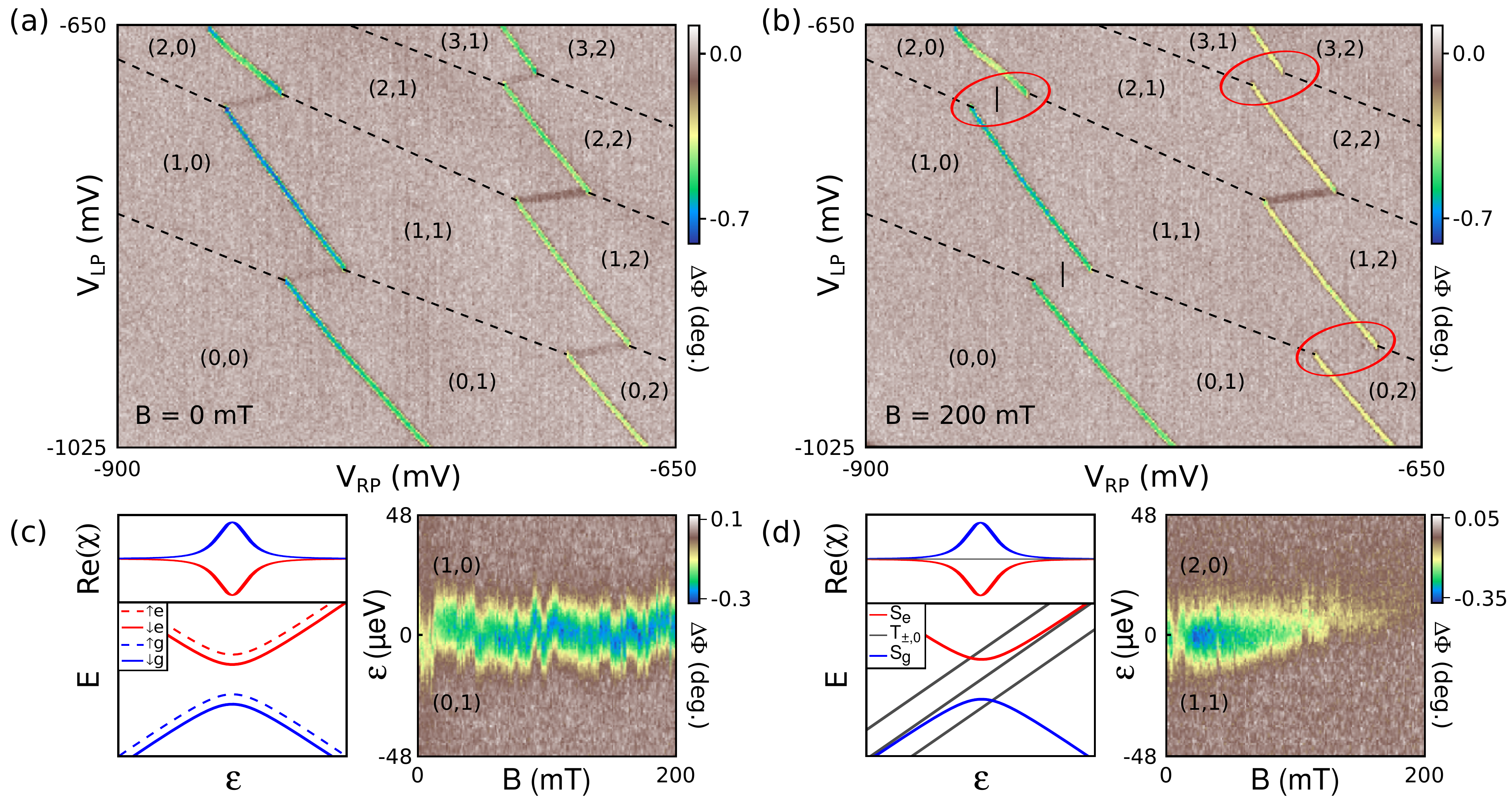}
\caption{\label{sense2} (Color online) The mesoscopic admittance of the DQD is sensitive to the total charge parity. (a) For $B$ = 0 mT, a phase response is detected at all interdot transitions. (b) At $B$ = 200 mT, the resonator is insensitive to charge dynamics at even parity charge transitions (circled in red). (c) Energy level diagram near the (1,0)$\leftrightarrow$(0,1) charge transition. Charge dynamics in the double dot lead to an effective ac susceptibility, $\chi$, which is maximal at $\epsilon$ = 0. The ac susceptibility is largely insensitive to magnetic field. (d) Energy level diagram near an even parity charge transition, e.g.\ (2,0)$\leftrightarrow$(1,1). Pauli exclusion prohibits triplet tunneling near $\epsilon$ = 0, however singlet state tunneling is allowed. At large magnetic fields, the $T_+$ triplet state becomes the ground state and the ac susceptibility is greatly reduced, quenching the phase response of the resonator.}
\end{center}	
\end{figure*}

Previous experiments probed spin-dependent effects in dc transport \cite{Hanson2007}. Here we examine the dynamical response of a DQD defined in a single InAs nanowire at rf frequencies of $\sim$ 0.5 GHz [Fig.\ 1(a)]. Samples are fabricated by dispersing InAs nanowires on an oxidized, high resistivity silicon substrate that is pre-patterned with local gate electrodes \cite{Fasth2005,Schroer2010,Nadj-Perge2010,Schroer2011}. The nanowire is separated from the gate electrodes by a 20 nm thick layer of SiN$_{\rm x}$ and electron beam lithography is used to define source and drain contacts to the wire. A scanning electron microscope (SEM) image of a finished sample is shown in Fig.\ \ref{sense1}(b). The sample is connected to a lumped element circuit consisting of a 110 nH surface mount inductor and its parasitic capacitance, resulting in a resonance frequency $f_{\rm c} = \omega_{\rm c}/2\pi \approx$ 560.9 MHz.

The sample is probed using rf-reflectometry \cite{Schoelkopf1998,Reilly2007,Cassidy2007}. Figure \ref{sense1}(c) displays the phase shift of the reflected signal, $\Delta \Phi$, as a function of left and right plunger gate voltages ($V_{\rm LP}$ and $V_{\rm RP}$), with no external magnetic field applied, $B$ = 0 T. The measured phase shift reflects the charge stability diagram of a DQD \cite{Wiel2002}, with charge stability islands labeled ($N_{\rm L}$, $N_{\rm R}$), where $N_{\rm L}$ ($N_{\rm R}$) are the number of electrons in the left (right) dot. When the DQD is deep in Coulomb blockade, the charge number is fixed, and the DQD is electrically decoupled from the resonator. At dot-lead and interdot charge transitions, the resonator's electric field induces charge dynamics, which in turn affect the amplitude and phase of the reflected rf signal \cite{Chorley2012}. We note that left dot charging transitions are much fainter due to a slower tunneling rate and a smaller capacitive coupling to the rf resonator \cite{Jung2012}. The absence of charge transitions in the lower left corner of Fig.\ \ref{sense1}(c) indicates that the DQD is emptied of all free electrons, resulting in a (0,0) charge state. We focus on rf spectroscopy of the device; dc transport measurements from similar devices have been presented elsewhere \cite{Nadj-Perge2010,Schroer2011,Jung2012}.

Radio frequency measurements of the charge stability diagram display a striking magnetic field dependence that is entirely absent in earlier experiments using a low frequency quantum point contact (QPC) charge detector \cite{Petta2005,Hanson2007}. Figure \ref{sense2}(a) shows the charge stability diagram, measured from the phase shift of the rf signal, for the first several charge transitions at $B$ = 0 T. Each interdot charge transition is visible, consistent with dc transport measurements on GaAs DQD devices \cite{Hanson2007}. In contrast, Fig.\ \ref{sense2}(b) shows the results from measurements taken with $B$ = 200 mT. Here the phase response is entirely suppressed at even parity interdot charge transitions (circled in red), while odd parity interdot charge transitions are essentially unaffected. The magnetic field dependence of the phase shift is measured as a function of detuning, $\epsilon$, at an odd parity transition in Fig.\ \ref{sense2}(c) and an even parity transition in Fig.\ \ref{sense2}(d) \cite{leverarm}. The phase shift is largely insensitive to field at the odd transition, while the signal at the even parity transition quickly diminishes in intensity with increasing magnetic field and also shifts to more positive values of detuning.

The charge parity dependence of this measurement may be explained considering the energy level diagram of a few-electron DQD \cite{Petta2005}.  At an odd parity interdot transition, the DQD forms a charge qubit, where the additional electron can occupy either the left or right dot. Figure \ref{sense2}(c) shows the energy eigenstates of a charge qubit as a function of detuning, $\epsilon$. Odd parity transitions are defined by two spin degenerate charge states which hybridize into ground (excited) eigenstates, denoted $g$ ($e$) in Fig.\ 2(c). Interdot tunnel coupling, $t_{\rm c}$, results in an energy splitting of $2 t_{\rm c}$ at $\epsilon$ = 0. When the thermal energy is less than the tunnel coupling, $k_{\rm B} T \ll 2 t_{\rm c}$, the ground state is occupied with unit probability. Here $k_{\rm B}$ is Boltzmann's constant. The resonator is sensitive to the charge susceptibility of the DQD, $\chi$, which is determined by the curvature of the energy levels \cite{Petersson2010}. A nonzero magnetic field lifts the Kramer's degeneracy, but does not change the charge susceptibility of the ground state, resulting in no magnetic field dependence of the interdot signal. The right panel of Fig.\ \ref{sense2}(c) displays the measured magnetic field dependence of the (1,0)$\leftrightarrow$(0,1) transition, showing that it is largely insensitive to magnetic field. The slight magnetic field dependence below 25 mT is attributed to a superconducting-normal transition in the solder and Al bond wires that form part of the resonator circuit.

In contrast, Fig.\ \ref{sense2}(d) shows the energy level diagram at an even parity transition. There are three spin triplet states ($T_+$, $T_-$, and $T_{\rm 0}$), in addition to ground and excited spin singlet states, denoted $S_{\rm g}$ and $S_{\rm e}$ \cite{Petta2005}. At $B$ = 0 and $\epsilon$ = 0, the ground state is a spin singlet, which has a non-zero charge susceptibility. Triplet tunneling at an interdot charge transition is forbidden due to the Pauli exclusion principle and the large exchange splitting of a doubly occupied quantum dot. As a result, the triplet energy levels are flat and have zero charge susceptibility. As the magnetic field is increased, the system's ground state transitions from a spin singlet to a spin triplet state, locking the system in a state with zero charge susceptibility.  This transition is thermally broadened, and there is a smooth decrease in the signal with increasing magnetic field, as shown in the right panel of Fig.\ \ref{sense2}(d).

A quantitative model for the response of the resonator may be developed using cavity input-output theory \cite{Walls1995,Clerk2010,Petersson2012}. The rf signal reflected by the device is given by:

\begin{equation}
  R = 1+\frac{i\kappa}{\Delta_{\rm c}/\hbar -i(\kappa+\kappa_{\rm i})/2+g_{\rm eff}\chi}
  \label{eq:oddreflect}
\end{equation}

where $\kappa$ is the cavity decay rate caused by coupling to the cavity port, $\kappa_{\rm i}$ is the decay rate due to internal loss mechanisms, and $\Delta_{\rm c} = h(f_{\rm c}-f_{\rm o})$ is the detuning of the resonator from the measurement frequency $f_{\rm o}$. The effective resonator -- charge coupling rate is $g_{\rm eff} = (2t_{\rm c}/\Omega) g_{\rm c}$, where $\Omega = \sqrt{4 t_{\rm c}^2 + \epsilon^2}$ and $g_{\rm c}$ is the bare coupling rate. The zero temperature charge susceptibility of the quantum dot is given by $\chi(T=0) = g_{\rm eff}/(i \Gamma/2 - \Delta/\hbar)$, where $\Delta = \Omega-h f_{\rm o}$, and $\Gamma$ is the charge relaxation rate. No term in Eqn.\ \ref{eq:oddreflect} depends on the magnetic field, confirming the magnetic field independence of odd parity transitions.

\begin{figure}[t]
\begin{center}
		\includegraphics[width=\columnwidth]{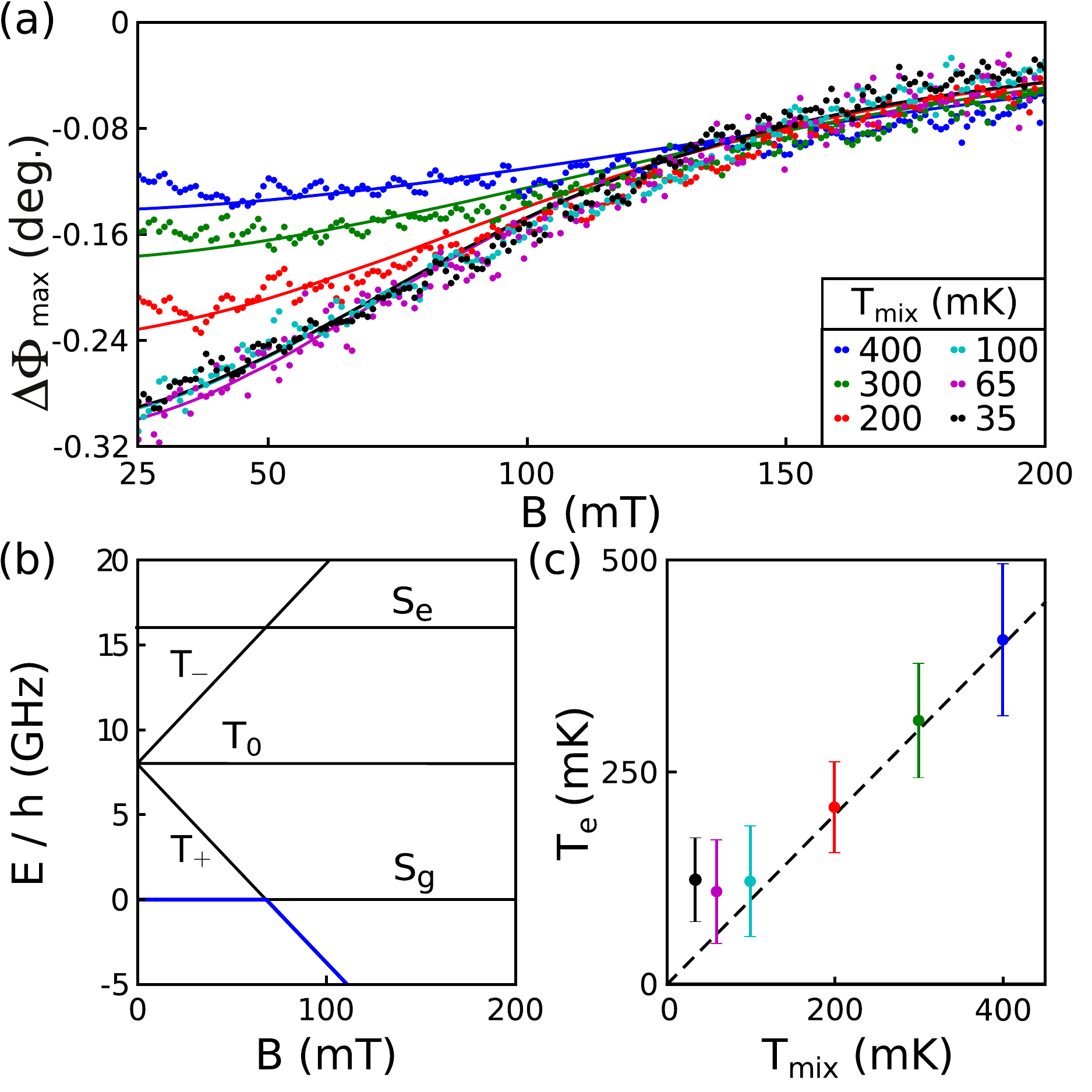}
\caption{\label{sense3} (Color online) Maximum phase shift, $\Delta \Phi_{\rm max}$, plotted as a function of magnetic field for various temperatures.  (a) Phase shifts extracted from data similar to Fig.\ \ref{sense2}(d), for temperatures from 35 to 400 mK.  Fits to Eqn.\ \ref{eq:spinreflection} are overlaid.  (b) Relevant energy levels plotted as a function of magnetic field; the ground state is outlined in blue.  A thermally broadened suppression of the phase shift occurs as T$_+$ becomes the system's ground state.  (c) Electron temperature, $T_{\rm e}$, extracted from the fits to the data in (a) as a function of the mixing chamber temperature, $T_{\rm mix}$. The electron temperature saturates at $\sim 130$ mK.}
\end{center}
\vspace{-0.4cm}
\end{figure}

At zero temperature and magnetic field, Eqn.\ \ref{eq:oddreflect} is equally valid for even and odd parity transitions. We may extend Eqn.\ \ref{eq:oddreflect} to the case of finite field and temperature by replacing the zero temperature value of $\chi$ with its thermal average. We calculate this quantity for the case of an even parity transition by identifying the charge susceptibility of the five relevant spin and charge states [see Fig.\ \ref{sense2}(d)], namely, $\chi(S_g) = -\chi(S_e) = \chi$ and $\chi(T_{\pm,0}) = 0$. With this, the thermally averaged susceptibility becomes

\begin{align}
  \left<\chi\right>&  =  \chi\left[\frac{e^{\Omega/2k_{\rm B}T_{\rm e}}-e^{-\Omega/2k_{\rm B}T_{\rm e}}}{Z(B,T_{\rm e})}\right]
\end{align}
where $Z(B,T_{\rm e})$ is the partition function over the five relevant states. The reflection coefficient is now given by:
\begin{equation}
 R(B,T_{\rm e}) = 1 + \frac{i\kappa}{\Delta_{\rm c}/\hbar -i(\kappa +\kappa_{\rm i})/2 + g_{\rm eff}\left<\chi\right>}.
 \label{eq:spinreflection}
\end{equation}

We reduce the number of free parameters in Eqn.\ \ref{eq:spinreflection} by first fitting the resonator response as a function of frequency, which determines $f_{\rm c}$ = 560.9 MHz, $\kappa/2\pi$ = 3.14 MHz, and $\kappa_i/2\pi$ = 1.02 MHz. Assuming $\Gamma/2\pi =$ 1 GHz from previous measurements \cite{Petta2004}, this leaves $g_{\rm c}$, $t_{\rm c}$ and $T_{\rm e}$ as the only unknown parameters.

We make quantitative comparisons with Eqn.\ \ref{eq:spinreflection} by measuring the maximal phase response as a function of magnetic field and temperature. We fit these data, which are shown in Fig.\ \ref{sense3}(a), by first convolving Eqn.\ \ref{eq:spinreflection} with a Gaussian of width 21 $\mu$eV \cite{Petersson2012} and then numerically computing the maximal phase shift. The convolution accounts for inhomogeneous broadening of the interdot charge transition due to charge fluctuations \cite{Hayashi2003,Petta2004}. For these fits, $g_{\rm c}$ and $t_{\rm c}$ are held constant across all data sets \cite{gfactor}. We extract a resonator -- charge coupling rate of $g_{\rm c}/2\pi$ = 16.7 MHz and the tunnel coupling $t_{\rm c}/h \sim$ 8.1 GHz. The resonator -- charge coupling rate is comparable with values observed in similar systems \cite{Frey2012,Petersson2012}. The electron temperature, $T_{\rm e}$, is also extracted for each value of the mixing chamber temperature, $T_{\rm mix}$, as shown in Fig.\ \ref{sense3}(c). The electron temperature tracks the mixing chamber temperature at high temperatures, and saturates around 130 mK, in agreement with other measurements on the same cryostat.

\begin{figure}
\begin{center}
		\includegraphics[width=\columnwidth]{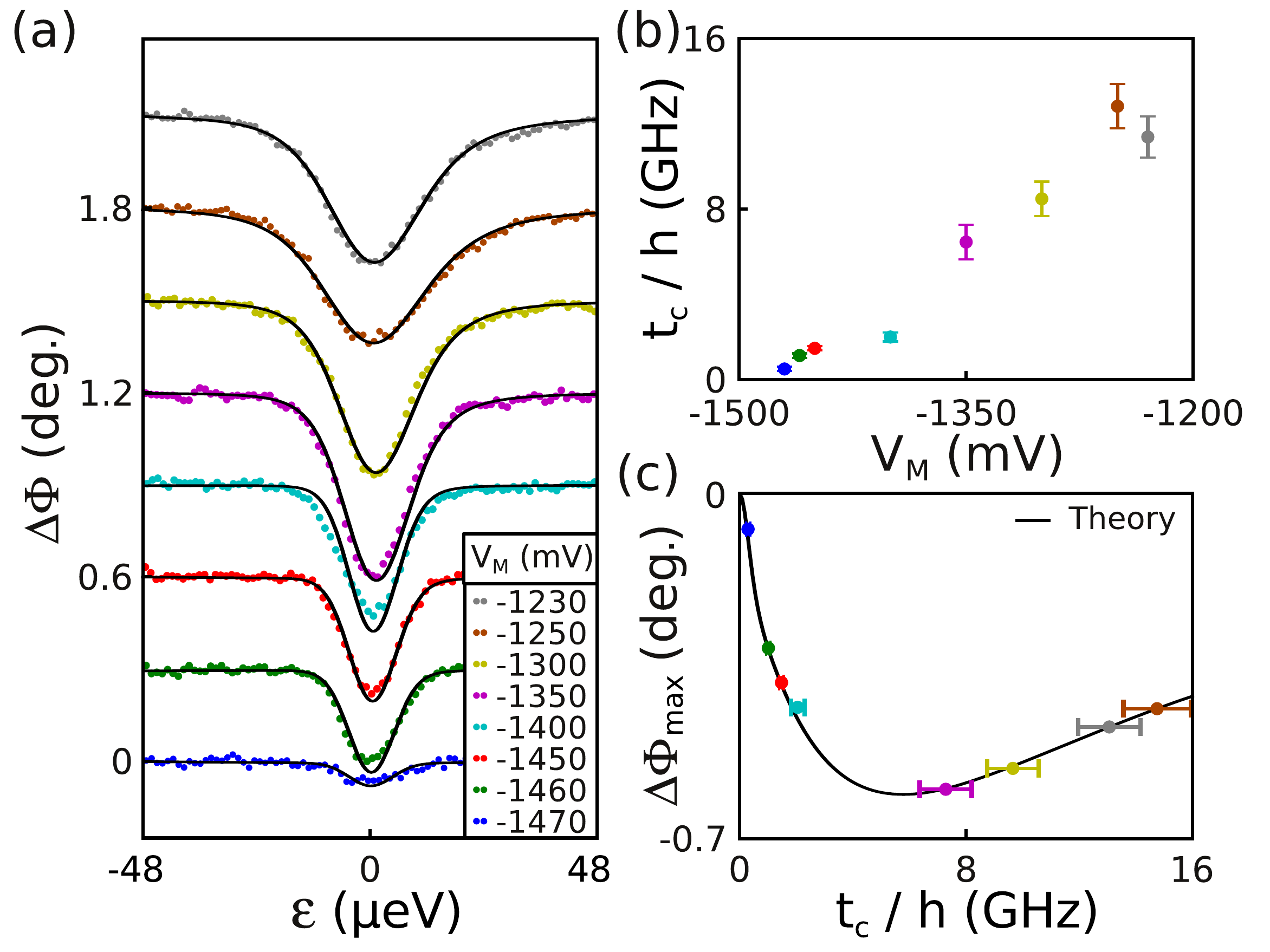}
\caption{\label{sense4} (Color online) Phase response as a function of interdot tunnel coupling. (a) Phase response at the (2,0)$\leftrightarrow$(1,1) charge transition for middle gate voltages ranging from -1230 to -1470 mV. Each trace has been vertically offset by 0.3$^\circ$ for clarity. (b)  The tunnel coupling extracted from the plots in (a).  No signal is observable below a middle gate voltage of -1470 mV. (c) The maximum phase shift measured for each curve in (a) as a function of the extracted tunnel coupling.  The solid line shows the theoretical response due to thermal mixing of the states at 130 mK; there is a strong suppression of the signal as the tunnel coupling approaches the thermal energy scale.}
\end{center}
\vspace{-0.4cm}
\end{figure}

We now focus on the $B$ = 0 phase response at even parity transitions. Figure \ref{sense4}(a) displays the observed phase response at the (2,0)$\leftrightarrow$(1,1) charge transition for different values of $V_{\rm M}$, which sets the interdot tunneling rate. Interdot tunnel coupling decreases as $V_{\rm M}$ is made more negative. We find that the measured phase shift first increases in amplitude and then vanishes rapidly. We fit these data using the procedure outlined above, with $g_{\rm c}$ = 16.7 MHz and $T_{\rm e}$ = 130 mK. The extracted values of $t_{\rm c}/h$ are shown in Fig.\ 4(b) and display a nearly monotonic dependence on $V_{\rm M}$, starting at 13 GHz for the most positive gate voltages and reaching 250 MHz for the most negative gate voltage. The ability to measure such small tunnel couplings highlights an advantage of this scheme over alternative methods that are limited to $t_{\rm c} \gg k_{\rm B} T$ \cite{Petta2004,DiCarlo_PRL_2004}.

In the zero temperature limit, the maximum phase response occurs as the tunnel coupling approaches the resonator frequency, $t_c/h \sim f_c$. At tunnel couplings much smaller than the resonator frequency, motion of the charge is highly nonadiabatic and the response is suppressed.  However, in contrast to studies involving high frequency superconducting cavities \cite{Frey2012,Petersson2012}, our resonator (at 560.9 MHz) represents an energy scale that is small compared to both the typical tunnel coupling and the thermal energy scale.  As such, the finite temperature plays a larger role in determining the thermally averaged interaction between the resonator and the sample than does the finite resonator frequency.  To illustrate this, Fig.\ \ref{sense4}(c) shows the maximum phase shift, $\Delta \Phi_{\rm max}$, extracted for each curve in Fig.\ \ref{sense4}(a). It is clear that the suppression of the resonator response takes place at $t_c/h$ $\approx$ 2--3 GHz, which is large compared to $f_{c}$. Predictions from the cavity input-output formalism are shown for comparison and agree well with the experimental data.

In conclusion, we have performed sensitive measurements of the charge state in an InAs nanowire DQD using rf-reflectometry on a coupled lumped element resonator. We can directly determine the total charge parity of the DQD at finite magnetic fields due to the different charge susceptibilities of singlet and triplet spin states. This measurement technique may be useful for determining which interdot charge transitions will exhibit Pauli blockade in dc transport, particularly in high effective mass systems, such as Si/SiGe quantum dots, where it is not always feasible to reach the (0,0) charge state. The magnetic field, temperature, and tunnel coupling dependence of the reflected rf signal agree well with cavity input-output theory, and gives a resonator-charge coupling rate of $g_{\rm c}/2\pi$ $\sim$ 17 MHz. Moreover, the phase response of the resonator can be used to extract the interdot tunnel coupling, even when the temperature is the dominant energy scale.

\begin{acknowledgments}
We acknowledge helpful discussions with Jacob Taylor. Research at Princeton was supported by Army Research Office grant W911NF-08-1-0189, DARPA QuEST grant HR0011-09-1-0007, and the NSF through the Princeton Center for Complex Materials (DMR-0819860) and CAREER program (DMR-0846341). Research was carried out in part at the Center for Functional Nanomaterials, Brookhaven National Laboratory, which is supported by DOE BES Contract No.\ DE-AC02-98CH10886. Partially sponsored by the United States Department of Defense. The views and conclusions contained in this document are those of the authors and should not be interpreted as representing the official policies, either expressly or implied, of the U.S. Government.
\end{acknowledgments}


\begin{thebibliography}{30}
\expandafter\ifx\csname natexlab\endcsname\relax\def\natexlab#1{#1}\fi
\expandafter\ifx\csname bibnamefont\endcsname\relax
  \def\bibnamefont#1{#1}\fi
\expandafter\ifx\csname bibfnamefont\endcsname\relax
  \def\bibfnamefont#1{#1}\fi
\expandafter\ifx\csname citenamefont\endcsname\relax
  \def\citenamefont#1{#1}\fi
\expandafter\ifx\csname url\endcsname\relax
  \def\url#1{\texttt{#1}}\fi
\expandafter\ifx\csname urlprefix\endcsname\relax\def\urlprefix{URL }\fi
\providecommand{\bibinfo}[2]{#2}
\providecommand{\eprint}[2][]{\url{#2}}

\bibitem[{\citenamefont{van~der Wiel et~al.}(2002)\citenamefont{van~der Wiel,
  Franceschi, Elzerman, Fujisawa, Tarucha, and Kouwenhoven}}]{Wiel2002}
\bibinfo{author}{\bibfnamefont{W.~G.} \bibnamefont{van~der Wiel}},
  \bibinfo{author}{\bibfnamefont{S.~D.} \bibnamefont{Franceschi}},
  \bibinfo{author}{\bibfnamefont{J.~M.} \bibnamefont{Elzerman}},
  \bibinfo{author}{\bibfnamefont{T.}~\bibnamefont{Fujisawa}},
  \bibinfo{author}{\bibfnamefont{S.}~\bibnamefont{Tarucha}}, \bibnamefont{and}
  \bibinfo{author}{\bibfnamefont{L.~P.} \bibnamefont{Kouwenhoven}}, \bibinfo{journal}{Rev. Mod. Phys.} \textbf{\bibinfo{volume}{75}}, \bibinfo{pages}{1} (\bibinfo{year}{2003}).

\bibitem[{\citenamefont{Oosterkamp et~al.}(1998)\citenamefont{Oosterkamp,
  Fujisawa, van~der Wiel, Ishibashi, Hijman, Tarucha, and
  Kouwenhoven}}]{Oosterkamp1998}
\bibinfo{author}{\bibfnamefont{T.~H.} \bibnamefont{Oosterkamp}},
  \bibinfo{author}{\bibfnamefont{T.}~\bibnamefont{Fujisawa}},
  \bibinfo{author}{\bibfnamefont{W.~G.} \bibnamefont{van~der Wiel}},
  \bibinfo{author}{\bibfnamefont{K.}~\bibnamefont{Ishibashi}},
  \bibinfo{author}{\bibfnamefont{R.~V.} \bibnamefont{Hijman}},
  \bibinfo{author}{\bibfnamefont{S.}~\bibnamefont{Tarucha}}, \bibnamefont{and}
  \bibinfo{author}{\bibfnamefont{L.~P.} \bibnamefont{Kouwenhoven}}, \bibinfo{journal}{Nature}
  \textbf{\bibinfo{volume}{395}}, \bibinfo{pages}{873} (\bibinfo{year}{1998}).

\bibitem[{\citenamefont{Ono et~al.}(2002)\citenamefont{Ono, Austing, Tokura,
  and Tarucha}}]{Ono2002}
\bibinfo{author}{\bibfnamefont{K.}~\bibnamefont{Ono}},
  \bibinfo{author}{\bibfnamefont{D.~G.} \bibnamefont{Austing}},
  \bibinfo{author}{\bibfnamefont{Y.}~\bibnamefont{Tokura}}, \bibnamefont{and}
  \bibinfo{author}{\bibfnamefont{S.}~\bibnamefont{Tarucha}}, \bibinfo{journal}{Science}
  \textbf{\bibinfo{volume}{297}}, \bibinfo{pages}{1313} (\bibinfo{year}{2002}).

\bibitem[{\citenamefont{Johnson et~al.}(2005)\citenamefont{Johnson, Petta,
  Marcus, Hanson, and Gossard}}]{Johnson2005a}
\bibinfo{author}{\bibfnamefont{A.~C.} \bibnamefont{Johnson}},
  \bibinfo{author}{\bibfnamefont{J.~R.} \bibnamefont{Petta}},
  \bibinfo{author}{\bibfnamefont{C.~M.} \bibnamefont{Marcus}},
  \bibinfo{author}{\bibfnamefont{M.~P.} \bibnamefont{Hanson}},
  \bibnamefont{and} \bibinfo{author}{\bibfnamefont{A.~C.}
  \bibnamefont{Gossard}}, \bibinfo{journal}{Phys. Rev. B} \textbf{\bibinfo{volume}{72}},
  \bibinfo{pages}{165308} (\bibinfo{year}{2005}).

\bibitem[{\citenamefont{Hayashi et~al.}(2003)\citenamefont{Hayashi, Fujisawa,
  Cheong, Jeong, and Hirayama}}]{Hayashi2003}
\bibinfo{author}{\bibfnamefont{T.}~\bibnamefont{Hayashi}},
  \bibinfo{author}{\bibfnamefont{T.}~\bibnamefont{Fujisawa}},
  \bibinfo{author}{\bibfnamefont{H.~D.} \bibnamefont{Cheong}},
  \bibinfo{author}{\bibfnamefont{Y.~H.} \bibnamefont{Jeong}}, \bibnamefont{and}
  \bibinfo{author}{\bibfnamefont{Y.}~\bibnamefont{Hirayama}},
  \bibinfo{journal}{Phys. Rev. Lett.} \textbf{\bibinfo{volume}{91}}, \bibinfo{pages}{226804} (\bibinfo{year}{2003}).

\bibitem[{\citenamefont{Petta et~al.}(2004)\citenamefont{Petta, Johnson,
  Marcus, Hanson, and Gossard}}]{Petta2004}
\bibinfo{author}{\bibfnamefont{J.~R.} \bibnamefont{Petta}},
  \bibinfo{author}{\bibfnamefont{A.~C.} \bibnamefont{Johnson}},
  \bibinfo{author}{\bibfnamefont{C.~M.} \bibnamefont{Marcus}},
  \bibinfo{author}{\bibfnamefont{M.~P.} \bibnamefont{Hanson}},
  \bibnamefont{and} \bibinfo{author}{\bibfnamefont{A.~C.}
  \bibnamefont{Gossard}}, \bibinfo{journal}{Phys. Rev. Lett.} \textbf{\bibinfo{volume}{93}},
  \bibinfo{pages}{186802}
  (\bibinfo{year}{2004}).

\bibitem[{\citenamefont{Petta et~al.}(2005)\citenamefont{Petta, Johnson,
  Taylor, Laird, Yacoby, Lukin, Marcus, Hanson, and Gossard}}]{Petta2005}
\bibinfo{author}{\bibfnamefont{J.~R.} \bibnamefont{Petta}},
  \bibinfo{author}{\bibfnamefont{A.~C.} \bibnamefont{Johnson}},
  \bibinfo{author}{\bibfnamefont{J.~M.} \bibnamefont{Taylor}},
  \bibinfo{author}{\bibfnamefont{E.~A.} \bibnamefont{Laird}},
  \bibinfo{author}{\bibfnamefont{A.}~\bibnamefont{Yacoby}},
  \bibinfo{author}{\bibfnamefont{M.~D.} \bibnamefont{Lukin}},
  \bibinfo{author}{\bibfnamefont{C.~M.} \bibnamefont{Marcus}},
  \bibinfo{author}{\bibfnamefont{M.~P.} \bibnamefont{Hanson}},
  \bibnamefont{and} \bibinfo{author}{\bibfnamefont{A.~C.}
  \bibnamefont{Gossard}}, \bibinfo{journal}{Science} \textbf{\bibinfo{volume}{309}}, \bibinfo{pages}{2180}
  (\bibinfo{year}{2005}).

\bibitem[{\citenamefont{Hanson et~al.}(2007)\citenamefont{Hanson, Kouwenhoven,
  Peta, Tarucha, and Vandersypen}}]{Hanson2007}
\bibinfo{author}{\bibfnamefont{R.}~\bibnamefont{Hanson}},
  \bibinfo{author}{\bibfnamefont{L.~P.} \bibnamefont{Kouwenhoven}},
  \bibinfo{author}{\bibfnamefont{J.~R.} \bibnamefont{Petta}},
  \bibinfo{author}{\bibfnamefont{S.}~\bibnamefont{Tarucha}}, \bibnamefont{and}
  \bibinfo{author}{\bibfnamefont{L.~M.~K.} \bibnamefont{Vandersypen}},
  \bibinfo{journal}{Rev. Mod. Phys.} \textbf{\bibinfo{volume}{79}}, \bibinfo{pages}{1217} (\bibinfo{year}{2007}).

\bibitem[{\citenamefont{Kouwenhoven et~al.}(1997)\citenamefont{Kouwenhoven,
  Oosterkamp, Danoesastro, Eto, Austing, Honda, and Tarucha}}]{Kouwenhoven1997}
\bibinfo{author}{\bibfnamefont{L.~P.} \bibnamefont{Kouwenhoven}},
  \bibinfo{author}{\bibfnamefont{T.~H.} \bibnamefont{Oosterkamp}},
  \bibinfo{author}{\bibfnamefont{M.~W.~S.} \bibnamefont{Danoesastro}},
  \bibinfo{author}{\bibfnamefont{M.}~\bibnamefont{Eto}},
  \bibinfo{author}{\bibfnamefont{D.~G.} \bibnamefont{Austing}},
  \bibinfo{author}{\bibfnamefont{T.}~\bibnamefont{Honda}}, \bibnamefont{and}
  \bibinfo{author}{\bibfnamefont{S.}~\bibnamefont{Tarucha}}, \bibinfo{journal}{Science}
  \textbf{\bibinfo{volume}{278}}, \bibinfo{pages}{1788} (\bibinfo{year}{1997}).

\bibitem[{\citenamefont{Buttiker et~al.}(1993)\citenamefont{Buttiker, Pretre,
  and Thomas}}]{Buttiker1993}
\bibinfo{author}{\bibfnamefont{M.}~\bibnamefont{Buttiker}},
  \bibinfo{author}{\bibfnamefont{A.}~\bibnamefont{Pretre}}, \bibnamefont{and}
  \bibinfo{author}{\bibfnamefont{H.}~\bibnamefont{Thomas}}, \bibinfo{journal}{Phys. Rev. Lett.}
  \textbf{\bibinfo{volume}{70}}, \bibinfo{pages}{4114} (\bibinfo{year}{1993}).

\bibitem[{\citenamefont{Pretre et~al.}(1996)\citenamefont{Pretre, Thomas, and
  Buttiker}}]{Pretre1996}
\bibinfo{author}{\bibfnamefont{A.}~\bibnamefont{Pretre}},
  \bibinfo{author}{\bibfnamefont{H.}~\bibnamefont{Thomas}}, \bibnamefont{and}
  \bibinfo{author}{\bibfnamefont{M.}~\bibnamefont{Buttiker}}, \bibinfo{journal}{Phys. Rev. B}
  \textbf{\bibinfo{volume}{54}}, \bibinfo{pages}{8130} (\bibinfo{year}{1996}).

\bibitem[{\citenamefont{Gabelli et~al.}(2006)\citenamefont{Gabelli, Feve,
  Berroir, Placais, Cavanna, Etienne, Jin, and Glattli}}]{Gabelli2006}
\bibinfo{author}{\bibfnamefont{J.}~\bibnamefont{Gabelli}},
  \bibinfo{author}{\bibfnamefont{G.}~\bibnamefont{Feve}},
  \bibinfo{author}{\bibfnamefont{J.~M.} \bibnamefont{Berroir}},
  \bibinfo{author}{\bibfnamefont{B.}~\bibnamefont{Placais}},
  \bibinfo{author}{\bibfnamefont{A.}~\bibnamefont{Cavanna}},
  \bibinfo{author}{\bibfnamefont{B.}~\bibnamefont{Etienne}},
  \bibinfo{author}{\bibfnamefont{Y.}~\bibnamefont{Jin}}, \bibnamefont{and}
  \bibinfo{author}{\bibfnamefont{D.~C.} \bibnamefont{Glattli}}, \bibinfo{journal}{Science} \textbf{\bibinfo{volume}{313}}, \bibinfo{pages}{5786} (\bibinfo{year}{2006}).

\bibitem[{\citenamefont{Mora and Le~Hur}(2010)}]{Mora2010}
\bibinfo{author}{\bibfnamefont{C.}~\bibnamefont{Mora}} \bibnamefont{and}
  \bibinfo{author}{\bibfnamefont{K.}~\bibnamefont{Le~Hur}}, \bibinfo{journal}{Nature Phys.}
  \textbf{\bibinfo{volume}{6}}, \bibinfo{pages}{697} (\bibinfo{year}{2010}).

\bibitem[{\citenamefont{Petersson et~al.}(2010)\citenamefont{Petersson, Smith,
  Anderson, Atkinson, Jones, and Ritchie}}]{Petersson2010}
\bibinfo{author}{\bibfnamefont{K.~D.} \bibnamefont{Petersson}},
  \bibinfo{author}{\bibfnamefont{C.~G.} \bibnamefont{Smith}},
  \bibinfo{author}{\bibfnamefont{D.}~\bibnamefont{Anderson}},
  \bibinfo{author}{\bibfnamefont{P.}~\bibnamefont{Atkinson}},
  \bibinfo{author}{\bibfnamefont{G.~A.~C.} \bibnamefont{Jones}},
  \bibnamefont{and} \bibinfo{author}{\bibfnamefont{D.~A.}
  \bibnamefont{Ritchie}},  \bibinfo{journal}{Nano Lett.} \textbf{\bibinfo{volume}{10}}, \bibinfo{pages}{2789}
  (\bibinfo{year}{2010}).

\bibitem[{\citenamefont{Cottet et~al.}(2011)\citenamefont{Cottet, Mora, and
  Kontos}}]{Cottet2011}
\bibinfo{author}{\bibfnamefont{A.}~\bibnamefont{Cottet}},
  \bibinfo{author}{\bibfnamefont{C.}~\bibnamefont{Mora}}, \bibnamefont{and}
  \bibinfo{author}{\bibfnamefont{T.}~\bibnamefont{Kontos}},
    \bibinfo{journal}{Phys. Rev. B}  \textbf{\bibinfo{volume}{83}}, \bibinfo{pages}{121311}  (\bibinfo{year}{2011}).

\bibitem[{\citenamefont{Chorley et~al.}(2012)\citenamefont{Chorley, Wabnig,
  Penfold-Fitch, Petersson, Frake, Smith, and Buitelaar}}]{Chorley2012}
\bibinfo{author}{\bibfnamefont{S.~J.} \bibnamefont{Chorley}},
  \bibinfo{author}{\bibfnamefont{J.}~\bibnamefont{Wabnig}},
  \bibinfo{author}{\bibfnamefont{Z.~V.} \bibnamefont{Penfold-Fitch}},
  \bibinfo{author}{\bibfnamefont{K.~D.} \bibnamefont{Petersson}},
  \bibinfo{author}{\bibfnamefont{J.}~\bibnamefont{Frake}},
  \bibinfo{author}{\bibfnamefont{C.~G.} \bibnamefont{Smith}}, \bibnamefont{and}
  \bibinfo{author}{\bibfnamefont{M.~R.} \bibnamefont{Buitelaar}},
  \bibinfo{journal}{Phys. Rev. Lett.}
  \textbf{\bibinfo{volume}{108}}, \bibinfo{pages}{036802}
  (\bibinfo{year}{2012}).

\bibitem[{\citenamefont{Wallraff et~al.}(2004)\citenamefont{Wallraff, Schuster,
  Blais, Frunzio, Huang, Majer, Kumar, Girvin, and Schoelkopf}}]{Wallraff2004}
\bibinfo{author}{\bibfnamefont{A.}~\bibnamefont{Wallraff}},
  \bibinfo{author}{\bibfnamefont{D.~I.} \bibnamefont{Schuster}},
  \bibinfo{author}{\bibfnamefont{A.}~\bibnamefont{Blais}},
  \bibinfo{author}{\bibfnamefont{L.}~\bibnamefont{Frunzio}},
  \bibinfo{author}{\bibfnamefont{R.~S.} \bibnamefont{Huang}},
  \bibinfo{author}{\bibfnamefont{J.}~\bibnamefont{Majer}},
  \bibinfo{author}{\bibfnamefont{S.}~\bibnamefont{Kumar}},
  \bibinfo{author}{\bibfnamefont{S.~M.} \bibnamefont{Girvin}},
  \bibnamefont{and} \bibinfo{author}{\bibfnamefont{R.~J.}
  \bibnamefont{Schoelkopf}}, \bibinfo{journal}{Nature} \textbf{\bibinfo{volume}{431}},
  \bibinfo{pages}{162} (\bibinfo{year}{2004}).

\bibitem[{\citenamefont{Delbecq et~al.}(2011)\citenamefont{Delbecq, Schmitt,
  Parmentier, Roch, Viennot, Feve, Huard, Mora, Cottet, and
  Kontos}}]{Delbecq2011}
\bibinfo{author}{\bibfnamefont{M.~R.} \bibnamefont{Delbecq}},
  \bibinfo{author}{\bibfnamefont{V.}~\bibnamefont{Schmitt}},
  \bibinfo{author}{\bibfnamefont{F.~D.} \bibnamefont{Parmentier}},
  \bibinfo{author}{\bibfnamefont{N.}~\bibnamefont{Roch}},
  \bibinfo{author}{\bibfnamefont{J.~J.} \bibnamefont{Viennot}},
  \bibinfo{author}{\bibfnamefont{G.}~\bibnamefont{Feve}},
  \bibinfo{author}{\bibfnamefont{B.}~\bibnamefont{Huard}},
  \bibinfo{author}{\bibfnamefont{C.}~\bibnamefont{Mora}},
  \bibinfo{author}{\bibfnamefont{A.}~\bibnamefont{Cottet}}, \bibnamefont{and}
  \bibinfo{author}{\bibfnamefont{T.}~\bibnamefont{Kontos}},
  \bibinfo{journal}{Phys. Rev. Lett.}
  \textbf{\bibinfo{volume}{107}},
  \bibinfo{pages}{256804}
  (\bibinfo{year}{2011}).

\bibitem[{\citenamefont{Frey et~al.}(2012)\citenamefont{Frey, Leek, Beck,
  Blais, Ihn, Ensslin, and Wallraff}}]{Frey2012}
\bibinfo{author}{\bibfnamefont{T.}~\bibnamefont{Frey}},
  \bibinfo{author}{\bibfnamefont{P.~J.} \bibnamefont{Leek}},
  \bibinfo{author}{\bibfnamefont{M.}~\bibnamefont{Beck}},
  \bibinfo{author}{\bibfnamefont{A.}~\bibnamefont{Blais}},
  \bibinfo{author}{\bibfnamefont{T.}~\bibnamefont{Ihn}},
  \bibinfo{author}{\bibfnamefont{K.}~\bibnamefont{Ensslin}}, \bibnamefont{and}
  \bibinfo{author}{\bibfnamefont{A.}~\bibnamefont{Wallraff}},
  \bibinfo{journal}{Phys. Rev. Lett.}
  \textbf{\bibinfo{volume}{108}}, \bibinfo{pages}{046807}
  (\bibinfo{year}{2012}).

\bibitem[{\citenamefont{Petersson et~al.}(2012)\citenamefont{Petersson, McFaul,
  Schroer, Jung, Taylor, Houck, and Petta}}]{Petersson2012}
\bibinfo{author}{\bibfnamefont{K.~D.} \bibnamefont{Petersson}},
  \bibinfo{author}{\bibfnamefont{L.~W.} \bibnamefont{McFaul}},
  \bibinfo{author}{\bibfnamefont{M.~D.} \bibnamefont{Schroer}},
  \bibinfo{author}{\bibfnamefont{M.}~\bibnamefont{Jung}},
  \bibinfo{author}{\bibfnamefont{J.~M.} \bibnamefont{Taylor}},
  \bibinfo{author}{\bibfnamefont{A.~A.} \bibnamefont{Houck}}, \bibnamefont{and}
  \bibinfo{author}{\bibfnamefont{J.~R.} \bibnamefont{Petta}},
  arXiv:1205.6767.

\bibitem[{\citenamefont{Petersson et~al.}(2012)\citenamefont{Petersson, McFaul,
  Schroer, Jung, Taylor, Houck, and Petta}}]{Jung2012}
   \bibinfo{author}{\bibfnamefont{M.}~\bibnamefont{Jung}},
    \bibinfo{author}{\bibfnamefont{M.~D.} \bibnamefont{Schroer}},
  \bibinfo{author}{\bibfnamefont{K.~D.} \bibnamefont{Petersson}},
  \bibnamefont{and}
  \bibinfo{author}{\bibfnamefont{J.~R.} \bibnamefont{Petta}},
  \bibinfo{journal}{Appl. Phys. Lett.} \textbf{\bibinfo{volume}{100}},
  \bibinfo{pages}{253508} (\bibinfo{year}{2012}).

\bibitem[{\citenamefont{Walls and Milburn}(1995)}]{Walls1995}
\bibinfo{author}{\bibfnamefont{D.~F.} \bibnamefont{Walls}} \bibnamefont{and}
  \bibinfo{author}{\bibfnamefont{G.~J.} \bibnamefont{Milburn}},
  \emph{\bibinfo{title}{Quantum Optics}} (\bibinfo{publisher}{Springer},
  \bibinfo{year}{1995}).

\bibitem[{\citenamefont{Clerk et~al.}(2010)\citenamefont{Clerk, Devoret,
  Girvin, Marquardt, and Schoelkopf}}]{Clerk2010}
\bibinfo{author}{\bibfnamefont{A.~A.} \bibnamefont{Clerk}},
  \bibinfo{author}{\bibfnamefont{M.~H.} \bibnamefont{Devoret}},
  \bibinfo{author}{\bibfnamefont{S.~M.} \bibnamefont{Girvin}},
  \bibinfo{author}{\bibfnamefont{F.}~\bibnamefont{Marquardt}},
  \bibnamefont{and} \bibinfo{author}{\bibfnamefont{R.~J.}
  \bibnamefont{Schoelkopf}}, \bibinfo{journal}{Rev. Mod. Phys.}
  \textbf{\bibinfo{volume}{82}}, \bibinfo{pages}{1155} (\bibinfo{year}{2010}).

\bibitem[{\citenamefont{Fasth et~al.}(2005)\citenamefont{Fasth, Fuhrer, Bjork,
  and Samuelson}}]{Fasth2005}
\bibinfo{author}{\bibfnamefont{C.}~\bibnamefont{Fasth}},
  \bibinfo{author}{\bibfnamefont{A.}~\bibnamefont{Fuhrer}},
  \bibinfo{author}{\bibfnamefont{M.~T.} \bibnamefont{Bjork}}, \bibnamefont{and}
  \bibinfo{author}{\bibfnamefont{L.}~\bibnamefont{Samuelson}},
   \bibinfo{journal}{Nano Lett.}
  \textbf{\bibinfo{volume}{5}}, \bibinfo{pages}{1487} (\bibinfo{year}{2005}).

\bibitem[{\citenamefont{Schroer et~al.}(2010)\citenamefont{Schroer, Xu,
  Bergman, and Petta}}]{Schroer2010}
\bibinfo{author}{\bibfnamefont{M.~D.} \bibnamefont{Schroer}},
  \bibinfo{author}{\bibfnamefont{S.~Y.} \bibnamefont{Xu}},
  \bibinfo{author}{\bibfnamefont{A.~M.} \bibnamefont{Bergman}},
  \bibnamefont{and} \bibinfo{author}{\bibfnamefont{J.~R.} \bibnamefont{Petta}},
   \bibinfo{journal}{Rev. Sci. Inst.}
  \textbf{\bibinfo{volume}{81}}, \bibinfo{pages}{023903}
  (\bibinfo{year}{2010}).

\bibitem[{\citenamefont{Nadj-Perge et~al.}(2010)\citenamefont{Nadj-Perge,
  Frolov, van Tilburg, Danon, Nazarov, Algra, Bakkers, and
  Kouwenhoven}}]{Nadj-Perge2010}
\bibinfo{author}{\bibfnamefont{S.}~\bibnamefont{Nadj-Perge}},
  \bibinfo{author}{\bibfnamefont{S.~M.} \bibnamefont{Frolov}},
  \bibinfo{author}{\bibfnamefont{J.~W.~W.} \bibnamefont{van Tilburg}},
  \bibinfo{author}{\bibfnamefont{J.}~\bibnamefont{Danon}},
  \bibinfo{author}{\bibfnamefont{Y.~V.} \bibnamefont{Nazarov}},
  \bibinfo{author}{\bibfnamefont{R.}~\bibnamefont{Algra}},
  \bibinfo{author}{\bibfnamefont{E.~P. A.~M.} \bibnamefont{Bakkers}},
  \bibnamefont{and} \bibinfo{author}{\bibfnamefont{L.~P.} \bibnamefont{Kouwenhoven}}, \bibinfo{journal}{Phys. Rev. B} \textbf{\bibinfo{volume}{81}},
  \bibinfo{pages}{201305} (\bibinfo{year}{2010}).

\bibitem[{\citenamefont{Schroer et~al.}(2011)\citenamefont{Schroer, Petersson,
  Jung, and Petta}}]{Schroer2011}
\bibinfo{author}{\bibfnamefont{M.~D.} \bibnamefont{Schroer}},
  \bibinfo{author}{\bibfnamefont{K.~D.} \bibnamefont{Petersson}},
  \bibinfo{author}{\bibfnamefont{M.}~\bibnamefont{Jung}}, \bibnamefont{and}
  \bibinfo{author}{\bibfnamefont{J.~R.} \bibnamefont{Petta}},
   \bibinfo{journal}{Phys. Rev. Lett.}
  \textbf{\bibinfo{volume}{107}}, \bibinfo{pages}{176811}
  (\bibinfo{year}{2011}).

\bibitem[{\citenamefont{Schoelkopf et~al.}(1998)\citenamefont{Schoelkopf,
  Wahlgren, Kozheznikov, Delsing, and Prober}}]{Schoelkopf1998}
\bibinfo{author}{\bibfnamefont{R.~J.} \bibnamefont{Schoelkopf}},
  \bibinfo{author}{\bibfnamefont{P.}~\bibnamefont{Wahlgren}},
  \bibinfo{author}{\bibfnamefont{A.~A.} \bibnamefont{Kozheznikov}},
  \bibinfo{author}{\bibfnamefont{P.}~\bibnamefont{Delsing}}, \bibnamefont{and}
  \bibinfo{author}{\bibfnamefont{D.~E.} \bibnamefont{Prober}},
   \bibinfo{journal}{Science}
  \textbf{\bibinfo{volume}{280}}, \bibinfo{pages}{1238} (\bibinfo{year}{1998}).

\bibitem[{\citenamefont{Reilly et~al.}(2007)\citenamefont{Reilly, Marcus,
  Hanson, and Gossard}}]{Reilly2007}
\bibinfo{author}{\bibfnamefont{D.~J.} \bibnamefont{Reilly}},
  \bibinfo{author}{\bibfnamefont{C.~M.} \bibnamefont{Marcus}},
  \bibinfo{author}{\bibfnamefont{M.~P.} \bibnamefont{Hanson}},
  \bibnamefont{and} \bibinfo{author}{\bibfnamefont{A.~C.}
  \bibnamefont{Gossard}},
   \bibinfo{journal}{Appl. Phys. Lett.}
  \textbf{\bibinfo{volume}{91}},
  \bibinfo{pages}{162101} (\bibinfo{year}{2007}).

\bibitem[{\citenamefont{Cassidy et~al.}(2007)\citenamefont{Cassidy, Dzurak,
  Clark, Petersson, Farrer, Ritchie, and Smith}}]{Cassidy2007}
\bibinfo{author}{\bibfnamefont{M.~C.} \bibnamefont{Cassidy}},
  \bibinfo{author}{\bibfnamefont{A.~S.} \bibnamefont{Dzurak}},
  \bibinfo{author}{\bibfnamefont{R.~G.} \bibnamefont{Clark}},
  \bibinfo{author}{\bibfnamefont{K.~D.} \bibnamefont{Petersson}},
  \bibinfo{author}{\bibfnamefont{I.}~\bibnamefont{Farrer}},
  \bibinfo{author}{\bibfnamefont{D.~A.} \bibnamefont{Ritchie}},
  \bibnamefont{and} \bibinfo{author}{\bibfnamefont{C.~G.} \bibnamefont{Smith}},
  \bibinfo{journal}{Appl. Phys. Lett.}
  \textbf{\bibinfo{volume}{91}}, \bibinfo{pages}{222104}
  (\bibinfo{year}{2007}).

\bibitem[{\citenamefont{Schroer et~al.}(2011)\citenamefont{Schroer, Petersson,
  Jung, and Petta}}]{leverarm}
\bibinfo{author}{\bibfnamefont{Raw data are measured as a function of $V_{\rm LP}$ along the vertical lines in Fig.\ 2(b). We convert from units of gate voltage to energy using the lever-arm $\alpha$ = 0.16 meV/mV}}.

\bibitem[{\citenamefont{Schroer et~al.}(2011)\citenamefont{Schroer, Petersson,
  Jung, and Petta}}]{gfactor}
\bibinfo{author}{\bibfnamefont{The g-factor is determined to be -8.4 from the electric-dipole spin resonance condition.}}

\bibitem{DiCarlo_PRL_2004} L. DiCarlo, H. J. Lynch, A. C. Johnson, L. I. Childress, K. Crockett, C. M. Marcus, M. P. Hanson, and A. C. Gossard, Phys. Rev. Lett. $\mathbf{92}$ 226801 (2004).

\end{thebibliography}
\end{document}